\documentclass[12pt]{article}
\usepackage{epsfig,amsfonts,amssymb}
\begin{document}

\voffset -0.7 true cm
\hoffset 1.1 true cm
\topmargin 0.0in
\evensidemargin 0.0in
\oddsidemargin 0.0in
\textheight 8.6in
\textwidth 7.1in
\parskip 10 pt

\newcommand{\be}{\begin{equation}}
\newcommand{\ee}{\end{equation}}
\newcommand{\bea}{\begin{eqnarray}}
\newcommand{\eea}{\end{eqnarray}}
\newcommand{\beas}{\begin{eqnarray*}}
\newcommand{\eeas}{\end{eqnarray*}}
\def\kl{{\frac{2 \pi l}{\beta}}}
\def\km{{\frac{2 \pi m}{\beta}}}
\def\kn{{\frac{2 \pi n}{\beta}}}
\def\kr{{\frac{2 \pi r}{\beta}}}
\def\ks{{\frac{2 \pi s}{\beta}}}
\def\b{{\beta}}
\font\cmsss=cmss8
\def\C{{\hbox{\cmsss C}}}
\font\cmss=cmss10
\def\bigC{{\hbox{\cmss C}}}
\def\scriptlap{{\kern1pt\vbox{\hrule height 0.8pt\hbox{\vrule width 0.8pt
  \hskip2pt\vbox{\vskip 4pt}\hskip 2pt\vrule width 0.4pt}\hrule height 0.4pt}
  \kern1pt}}
\def\ba{{\bar{a}}}
\def\bb{{\bar{b}}}
\def\bc{{\bar{c}}}

\def\Bigggl{\mathopen\Biggg}
\def\Bigggr{\mathclose\Biggg}
\def\Biggg#1{{\hbox{$\left#1\vbox to 25pt{}\right.\n@space$}}}
\def\n@space{\nulldelimiterspace=0pt \m@th}
\def\m@th{\mathsurround = 0pt}

\begin{titlepage}
\begin{flushright}
{\small CU-TP-1074} \\
{\small hep-th/0211138}
\end{flushright}

\begin{center}

\vspace{2mm}

{\Large \bf Supergravity, Supermembrane and M(atrix) model on PP-Waves}

\vspace{3mm}

Norihiro Iizuka

\vspace{1mm}

{\small \sl Department of Physics} \\
{\small \sl Columbia University, New York, NY 10027} \\
{\small \tt iizuka@phys.columbia.edu}
\vspace{1mm}

\end{center}

\vskip 0.3 cm

\noindent
In the first part of this paper, 
we study the back-reaction of large-N light cone momentum 
on the maximally supersymmetric anti-pp-wave background. 
This gives the type IIA geometry of large-N D0-branes on curved space with fluxes. 
By taking an appropriate decoupling limit, we conjecture a new duality between 
string theory on that background and dual field theory on D0-branes which 
we derive by calculating linear coupling terms. Agreement of decoupling quantities, 
$SO\left(3\right) \times SO\left(6\right)$ isometry and Higgs branch on both 
theories are shown. Also we find whenever dual field theory is weakly coupled, 
the curvature of the geometry is large. 
In the second part of this paper, 
we derive the supermembrane action on a general pp-wave background only through the 
properties of null Killing vector and 
through this, derive the Matrix model.

\end{titlepage}

%%%%%%%%%%%%%%%%%%%%%%%%%%%%%%%%%%%%%%%%%%%%%%%%%%%%%%%%%%%%%%%%%%%%%%%%%%%%%%
\section{Introduction}
%%%%%%%%%%%%%%%%%%%%%%%%%%%%%%%%%%%%%%%%%%%%%%%%%%%%%%%%%%%%%%%%%%%%%%%%%%%%%%
At the present time, two concrete examples of the holographic principle 
are well known. 
One is AdS/CFT and the other is the M(atrix) model of 
M-theory \cite{Reviews}. 
The latter is conjectured to be the microscopic definition of 11 
dimensional M-theory in the DLCQ description and its action is described 
by the large-N limit of dimensional reduction of 16 supercharges $U(N)$ gauge theory to 
$0+1$ dimensions \cite{Halpern}:
\label{BFSS}
\begin{eqnarray}
S_0 = \int d\tau 
Tr \left(
\sum_{i=1}^9 {1 \over 2} \dot{X}^{i2} 
+ {1 \over 4} \sum_{i,j=1}^9 \left[X^i , X^j \right]^2 
+ i \Psi^T \dot{\Psi} 
- \sum_{i=1}^9 \Psi^T \gamma^i \left[ X^i , \Psi \right] 
\right)
\end{eqnarray}

Interest in this model was revived after the paper \cite{BMN} where the Matrix model on
maximally supersymmetric pp-wave is derived containing extra mass terms: 
\label{mass}
\begin{eqnarray}
S_{mass} = \int d\tau
Tr \left(
- \sum_{i=1}^3 {1 \over 2} \left({ \mu \over 3} \right)^2 X^{i2} 
- \sum_{i=4}^9 {1 \over 2} \left({ \mu \over 6} \right)^2 X^{i2} 
+ {i \mu \over 3} \sum_{i,j,k=1}^3 \epsilon_{ijk} X^i  X^j  X^k 
- {i \mu \over 4} \Psi^T \gamma_{123} \Psi 
\right)
\end{eqnarray}

This corresponds to a background that is the maximally supersymmetric pp-wave metric \cite{pp,HDp},
\be
\label{ppmetric}
ds^2 = 2 dx^{+} dx^{-} 
- \sum_{i=1}^3 \left({\mu \over 3} \right)^2 x^{i2} dx^{+2} - \sum_{i=4}^9 \left({\mu \over 6} \right)^2 x^{i2} dx^{+2}
+ \sum_{i=1}^9 dx^{i2} \,.
\ee
\[ 
F^{\left(4 \right)} = \mu dx^{+} \wedge dx^1 \wedge dx^2 \wedge dx^3  
\]

Several papers about analysis and generalization of this action soon 
followed \cite{DSR,CLP2,DBak,Bonelli,SugiYoshi,spectrum1,spectrum2,SUSYalgebra,KLY,objects,Lee,mark,NKim}. 
A big difference of this model 
from one on flat space is the non-existence of flat direction 
due to the mass terms which are induced by pp-wave geometry. This improves the situation, 
since we do not have to worry about the threshold bound state, which is 
a very difficult problem in the flat space case.

On the other hand, 
it is well-known that the M(atrix) theory action is the same action of type 
IIA large-N D0-branes action in flat space background. 
This relation is clarified by the paper \cite{Sen, Seiberg}
and its claim that by infinitely boosting a large-N IIA D0 system,
 the small spacelike compactified radius along $x^{10}$ is infinitely 
boosted to a large lightlike radius along $x^{-}$. 
The reason we need to take large-N limit is 
because we want to take lightlike circle 
to infinity while keeping $p_{-}$ fixed. 
A similar story should hold in the M(atrix) model on the pp-wave background. 
In \cite{DSR}, IIA background metric on which 
D0-branes action is identical to BMN action is explicitly derived. 
They derived it 
by ``unboosting '' 11 dimensional pp-wave metric and then 
compactifying this to 10 dimensions. The metric is written in string frame as,
\be
\label{DSRmetric}
ds^2 = g_{00} dt^2 + g_{AB} dx^{A} dx^{B} \,.
\ee
\[
g_{00} = -e^{-{2 \over 3}\phi}, g_{AB} = \delta_{AB} e^{{2 \over 3} \phi}
\]
\[
e^{{4 \over 3} \phi} = 1 - {1 \over 4}F^2
\]
\[
A_0 = {- {F^2 \over 4} \over {1 - {F^2 \over 4}}}
\]
\[
F^2 \equiv  \left( {\mu \over 3} \right)^2 
\left( x_{1}^2 + x_{2}^2 + x_{3}^2  \right) 
+ \left( {\mu \over 6} \right)^2 
\left( x_{4}^2 + \cdots + x_{9}^2 \right)
\]
\[
F_{0123} = {\mu \over 2} { {1 - {F^2 \over 2}} \over { 1 - {F^2 \over 4}}} , 
H_{123} =  -{\mu \over 2}
\]
The above metric makes sense only when curvature is small, that is,
 only when $F^2 \ll 4$. Actually by using this property 
\cite{TaylorRaamsdonk}, 
we can construct the D0-branes action in this weak background and it is 
the same action of BMN \cite{DSR}.

In \cite{IMSY} it was shown that the back-reaction 
of the large-N D0-brane charge on the geometry is important. 
In the flat space background case, we know large-N D0-branes action 
is dual to string theory on a warped $AdS_2 \times S^8$ background, 
and depending on the energy scale we consider, 
we have different description of this 0+1 dimensional theory. 
The phase diagram of this D0-branes system is shown in the beautiful paper
 \cite{IMSY}.

Motivated by these developments, it would be interesting to 
establish a similar correspondence in the non-trivial background 
(\ref{DSRmetric}), that is, 
to obtain string theory on the near-horizon geometry of large-N D0-branes in 
the background geometry (\ref{DSRmetric}) that is dual to BMN action. 
Unfortunately we don't know the 11-dimensional supergravity solution 
which corresponds to D0-branes on geometry (\ref{DSRmetric}) 
in 10-dimension. 
But instead, we know anti-D0-branes 
(=graviton which propagate along $- x^{10}$ direction) solution on geometry (\ref{DSRmetric}). 
Equivalently we know D0-branes on all RR fields sign 
flipped geometry of (\ref{DSRmetric}), 
which corresponds to ``anti'' pp-wave in 
11-dimension. Therefore we study the back-reaction of large-N D0-brane 
charge on anti-pp-wave background in this paper\footnote{The author is 
grateful to Juan Maldacena for pointing out 
a mis-interpretation of anti-D0-branes as D0-branes in the first 
version of this paper.}.

The organization of this paper is composed of two parts. 
In section 2, we study the effect of the back-reaction of large-N 
D0-brane charge on the maximally supersymmetric anti-pp-wave. 
By taking an appropriate decoupling limit, we derive the IIA 
geometry which is given by eq. (\ref{decouplingmetric}) with flux 
(\ref{field}). 
Then we consider the dual field theory whose action is 
derived by calculating the linear coupling terms of D0-branes 
with background (\ref{antiDSRmetric}). We find agreement 
about well-defined ${\cal{O}}\left( \alpha'^2 \mu \right)$ deformation, 
$SO\left(3\right) \times SO\left(6\right)$ isometry and Higgs branch on both theories. 
Also it is shown that whenever the dual field theory can be described by 
perturbation theory, the curvature of the geometry (\ref{decouplingmetric}) is large. 
In section 3, this is completely independent work to 
section 2, we try to derive the Matrix model on a 
general pp-wave background. We consider the supermembrane action on a 
general pp-wave background up to quadratic terms of fermion fields. 
Then we prove that there are no higher order corrections only through the 
properties of null Killing vector. Finally, 
we obtain the Matrix model on a general pp-wave background 
and we study its supersymmetry 
which has the expected form given the background. In the appendix, 
we summarize the relevant facts about general pp-waves.

%%%%%%%%%%%%%%%%%%%%%%%%%%%%%%%%%%%%%%%%%%%%%%%%%%%%%%%%%%%%%%%%%%%%%%%%%
\section{Back-reaction of large-N D0-brane charge on anti-pp-wave}
%%%%%%%%%%%%%%%%%%%%%%%%%%%%%%%%%%%%%%%%%%%%%%%%%%%%%%%%%%%%%%%%%%%%%%%%%

%%%%%%%%%%%%%%%%%%%%%%%%%%%%%%%%%%%%%%%%%%%%%%%%%%%%%%%%%%%%%%%%%%%%%%%%%%
\subsection{Type IIA background}
%%%%%%%%%%%%%%%%%%%%%%%%%%%%%%%%%%%%%%%%%%%%%%%%%%%%%%%%%%%%%%%%%%%%%%%%%%
The gravitational wave with N units of momentum in flat 
11 dimensional space 
has the supergravity background 
\be
\label{gwavemetric}
ds^2 = 2 dx^{+} dx^{-} + \left( {C \over | \vec{x} | ^7} \right)
dx^{-2} + d \vec{x}^2 \,. 
\ee
\[
F^{\left( 4 \right)} = 0
\]
\[
C \equiv 2^8 \pi^{9 \over 2} \Gamma \left({7 \over 2}\right) \alpha'^5
 \left( g_{YM}^2 N \right)
\]
$F^{\left( 4 \right)}$ is field strength of 3-form gauge potential and 
the constant $C$ is checked from comparison of this metric when KK reduced
 to the 10 dimensional metric describing N D0-branes.
Now because of enough supersymmetry, the superposition of ``anti''-pp-wave metric 
\be
\label{ppmetric}
ds^2 = 2 dx^{+} dx^{-} 
- \sum_{i=1}^3 \left({\mu \over 3} \right)^2 x^{i2} dx^{-2} 
- \sum_{i=4}^9 \left({\mu \over 6} \right)^2 x^{i2} dx^{-2}
+ \sum_{i=1}^9 dx^{i2} 
\ee
\[ 
F^{\left(4 \right)} = \mu dx^{-} \wedge dx^1 \wedge dx^2 \wedge dx^3  
\]
and gravitational wave (\ref{gwavemetric}) is also 
a solution of Einstein equations. 
We combine them using the 
ansatz of plane wave,
\[
ds^2 = 2 dx^{+} dx^{-} + H \left(\vec{x} \right) dx^{-2} + d \vec{x}^2
\]
\be
\label{H}
H \equiv \left( {C \over | \vec{x} | ^7} \right) 
- F^2 = \left( {C \over | \vec{x} | ^7} \right) 
-  \left[ \left( {\mu \over 3} \right)^2 
\left( x_{1}^2 + x_{2}^2 + x_{3}^2  \right) 
+ \left( {\mu \over 6} \right)^2 
\left( x_{4}^2 + \cdots + x_{9}^2 \right) \right]
\ee
\[
F_{-123} = \mu
\]
where $F^2$ is defined in eq. (\ref{DSRmetric}). 
The only nontrivial component of the Ricci tensor is 
$R_{--} \sim \partial_i^2 H \sim
 -\mu^2$ (except at $\vec{x} = 0$). Because 
the curvature scalar vanishes, the 
superimposed solution satisfies the 
Einstein equation $R_{--} \sim -F_{-ijk} F_{-}^{ijk} \sim -\mu^2$ as expected.
Note that the solution (\ref{H}) preserves one half of the supersymmetry.
Knowing the 11 dimensional metric, it is straightforward to derive the 
10 dimensional metric in string frame. Taking into account $\mu \to {\mu \over \sqrt{2}}$ by unboosting, 
the metric is,
\be
\label{superimposedmetric}
ds^2 = g_{00} dt^2 + g_{AB} dx^{A} dx^{B} \,.
\ee
\[
g_{00} = -e^{-{2 \over 3}\phi}, g_{AB} = \delta_{AB} e^{{2 \over 3} \phi}
\]
\[
e^{{4 \over 3} \phi} = 1 + {1 \over 4}H
\]
\[
A_0 = { - {H \over 4} \over {1 + {H \over 4}}}
\]
\[ 
F_{0123} = - {\mu \over 2} {{1 + {H \over 2}} \over {1 + {H \over 4} } },
H_{123} = - {\mu \over 2} 
\]
This is the IIA supergravity solution 
called homogeneous D0-branes (HD0-branes) in \cite{HDp}.
Note that the above solution describing all N D0-branes sit at the center of the space, but 
we can put the center of each harmonic function anywhere, this is easily seen in 11-dimensional 
solution (\ref{H}) because of supersymmetry. 
Physically geometry is the same as considering usual pp-wave with N anti-D0-branes. The superposition of 
N D0-branes and usual pp-wave is not a 11 dimensional supergravity solution unfortunately.
%%%%%%%%%%%%%%%%%%%%%%%%%%%%%%%%%%%%%%%%%%%%%%%%%%%%%%%%%%%%%%%%%%%%%%%%%%
\subsection{Decoupling limit and near horizon geometry}
%%%%%%%%%%%%%%%%%%%%%%%%%%%%%%%%%%%%%%%%%%%%%%%%%%%%%%%%%%%%%%%%%%%%%%%%%%
As usual in AdS/CFT duality, we must take the near-horizon geometry of 
the metric (\ref{superimposedmetric}) so that the boundary theory decouples 
from the bulk theory. The decoupling limit we take is
\be
\label{decouplinglimit}
U^2 \equiv {1 \over \alpha'^2} \sum_{i=1}^{3} x^{i2} = \mbox{fixed}, \hspace{2mm} 
V^2 \equiv {1 \over \alpha'^2} \sum_{i=4}^{9} x^{i2} = \mbox{fixed},
\ee
\[
g_{YM}^2 = {1 \over 4 \pi^2} {g_s \over \alpha'^{3 \over 2}} = \mbox{fixed},
\hspace{2mm} 
\mu \alpha'^2 = \mbox{fixed}, \hspace{2mm} 
\alpha' \to 0
\]

The type IIA supergravity solution in this decoupling limit is
\be
\label{decouplingmetric}
ds^2 = \alpha' 
\left[- g_0^{- {1 \over 2}} dt^2 + g_0^{1 \over 2} \left(
dU^2 + U^2 d \Omega_2^2 \right) + g_0^{1 \over 2} \left(
dV^2 + V^2 d \Omega_5^2 \right) \right]
\ee
where $g_0$ is well-defined:
\be
\label{g_0def}
g_0 = {1 \over 4} \left[ {2^8 \pi^{9 \over 2} \Gamma\left({7 \over 2}\right) \left(g_{YM}^2 N \right) 
\over \left(U^2 + V^2\right)^{7 \over 2}} 
- \left( {\alpha'^2 \mu \over 3}\right)^2 U^2 
- \left( {\alpha'^2 \mu \over 6}\right)^2 V^2 \right]
\ee

In this decoupling limit, the NSNS and RR field strengths are
\be
\label{field}
H_{123} = - \alpha' \left({\alpha'^2 \mu \over 2}\right), \hspace{2mm}
F_{0123} = - \alpha' \left({ \alpha'^2 \mu }\right)
\ee

You can easily see that since both the metric and the NSNS field strength are ${\cal{O}} \left(\alpha' \right)$, 
their contributions are at the same order and 
give rise to the well-defined Nambu-Goto action, 
that is, $\alpha'$ cancel out in 
$S_{NG} \sim {1 \over \alpha'} \int d^2\sigma \sqrt{ det \left( G+B \right)} $. 
And of course these fluxes are supported by a curvature term which is proportional to $\alpha'^2 \mu$ in $g_0$\footnote{The author thanks Koenraad Schalm for a useful discussion on this point.}. 
Now it is clear that this background has $SO(9) \to SO(3) \times SO(6)$ symmetry breaking because of the ${\cal{O}}\left( \alpha'^3 \mu \right)$ terms in the flux and the metric. 
Here we take parameter $\mu$ as (\ref{decouplinglimit}). 
If we take different limit of $\mu$, the results are very different. 
For example if we take $\mu = \mbox{fixed}$, 
the effect of $\mu$ is negligible. The geometry is warped $AdS_2 \times S^8$ space and the rotational symmetry is enhanced to $SO(9)$. 
If we take a limit where $\mu$ is bigger than ${\cal{O}} \left(\alpha'^{-2} \right)$, the  
IIA metric doesn't make sense since the curvature becomes singular 
at $g_0 = 0$, i.e. from 11 dimensional point of view, 
the sign of metric in $ \left( dx^{10} \right)^2$ changes and becomes 
timelike, therefore compactification doesn't make sense at this point.

The curvature of this metric is (for simplicity taking $U$ and $V$ to be the same order)
\be
\label{curvature}
\alpha' {\cal{R}} \sim g_0^{-{1 \over 2}} U^{-2} 
\ee
So the curvature small condition is satisfied for 
\be
\label{U_mu>U_gy}
U \ll \left(g_{YM}^2 N \right)^{1 \over 3} \hspace{3mm} \mbox{if \hspace{1mm} $\left(g_{YM}^2 N \right)^{1 \over 3} \lesssim \left(\alpha'^2 \mu \right)^{-{1 \over 3}}$}
\ee
\be
\label{U_mu<U_gy}
U \ll \left(g_{YM}^2 N \right)^{1 \over 9} \left(\alpha'^2 \mu \right)^{-{2 \over 9}} \hspace{3mm} \mbox{if \hspace{1mm} $\left(g_{YM}^2 N \right)^{1 \over 3} \gtrsim \left(\alpha'^2 \mu \right)^{-{1 \over 3}}$}
\ee
Also the dilaton small condition is satisfied for 
\be
\label{dilaton}
{\left(g_{YM}^2 N \right)^{1 \over 3} \over N^{4 \over 21}} \ll U \lesssim \left( g_{YM}^2 N \right)^{-{2 \over 3}} \left( \alpha'^2 \mu \right)^{-1} N^{2 \over 3}
\ee
Note that the second inequality in eq. (\ref{dilaton}) is weaker than curvature small 
condition in eq. (\ref{U_mu>U_gy}) and (\ref{U_mu<U_gy}). 
We claim that IIA string theory on this background 
(\ref{decouplingmetric}) with fluxes (\ref{field}) 
is well-defined as long as we consider theory at energy scale region $U$:
\be
\label{case1U}
{\left(g_{YM}^2 N \right)^{1 \over 3} \over N^{4 \over 21}} \ll U \ll \left(g_{YM}^2 N \right)^{1 \over 3} \hspace{3mm} \mbox{if \hspace{1mm} $\left(g_{YM}^2 N \right)^{1 \over 3} \lesssim \left(\alpha'^2 \mu \right)^{-{1 \over 3}}$}
\ee
\be
\label{case2U}
{\left(g_{YM}^2 N \right)^{1 \over 3} \over N^{4 \over 21}} \ll U \ll \left(g_{YM}^2 N \right)^{1 \over 9} \left(\alpha'^2 \mu \right)^{-{2 \over 9}} \hspace{3mm} \mbox{if \hspace{1mm} $\left(g_{YM}^2 N \right)^{1 \over 3} \gtrsim \left(\alpha'^2 \mu \right)^{-{1 \over 3}}$}
\ee
%%%%%%%%%%%%%%%%%%%%%%%%%%%%%%%%%%%%%%%%%%%%%%%%%%%%%%%%%%%%%%%%%%%%%%%%%%
\subsection{Dual field theory}
%%%%%%%%%%%%%%%%%%%%%%%%%%%%%%%%%%%%%%%%%%%%%%%%%%%%%%%%%%%%%%%%%%%%%%%%%%
The theory on D0-branes is derived by assuming the linear coupling of D0-branes 
with these background field:
\be
\label{antiDSRmetric}
ds^2 = g_{00} dt^2 + g_{AB} dx^{A} dx^{B} \,.
\ee
\[
g_{00} = -e^{-{2 \over 3}\phi}, g_{AB} = \delta_{AB} e^{{2 \over 3} \phi}
\]
\[
e^{{4 \over 3} \phi} = 1 - {1 \over 4} F^2
\]
\[
A_0 = {{F^2 \over 4} \over {1 - {F^2 \over 4}}}
\]
\[
F^2 \equiv  \left( {\mu \over 3} \right)^2 
\left( x_{1}^2 + x_{2}^2 + x_{3}^2  \right) 
+ \left( {\mu \over 6} \right)^2 
\left( x_{4}^2 + \cdots + x_{9}^2 \right)
\]
\[
F_{0123} = -{\mu \over 2} {{4 - 2 F^2} \over {4 - F^2}} , 
H_{123} =  -{\mu \over 2}
\]
by following the methods of \cite{TaylorRaamsdonk}.
Note that this metric is derived by compactifying the ``anti''-pp-wave 
geometry along $x^{10}$ direction and taking into account the rescaling of 
$\mu \to {\mu \over \sqrt{2}}$, therefore the sign of RR-field $A_0$ and 
$F_{0123}$ is flipped compared with (\ref{DSRmetric}). 
By taking the weak curvature limit and neglecting the ${\cal{O}}(\left(F^2\right)^2)$ terms, the above IIA background is 
approximated as 
\begin{eqnarray*}
\label{weakBG}
\phi \sim - {3 \over 16} F^2, \hspace{2mm} h_{00} \sim - {1 \over 8} F^2, 
\hspace{2mm} h_{AB} \sim - {1 \over 8} F^2 \delta_{AB}, \\
A_0 \sim {F^2 \over 4}, \hspace{2mm} C_{0ij} \sim \epsilon_{ijk} {\mu \over 6} x^k, 
\hspace{2mm} B_{ij} \sim - \epsilon_{ijk} {\mu \over 6} x^k
\end{eqnarray*}
The dual field theory has action\footnote{We use the notation of \cite{TaylorRaamsdonk}}:
\[
S = S_{flat} + S_{linear}
\]
\begin{eqnarray*}
S_{linear} &=& \int dt \Bigl[ {1 \over 4} \partial_A \partial_B  h_{\mu \nu} I^{\mu \nu \left( A B \right)}_h 
+ {1 \over 2} \partial_A \partial_B \phi I^{\left( A B \right)}_{\phi} \\
& & + {1 \over 2} \partial_A \partial_B 
A_{\mu} I^{\mu \left( A B \right)}_{0} 
+ \partial_A  B_{\mu \nu} I^{\mu \nu \left( A \right)}_{s} + \partial_A  C_{\mu \nu \lambda} 
I^{\mu \nu \lambda \left( A \right)}_{2} \Bigr] \\
&=& \int dt \Bigl[ - {1 \over 16} \partial_A \partial_B F^2 \left( T^{--\left(AB\right)} + {\cal{O}}(\alpha'^{11 \over 2})\right) 
+ \epsilon_{ijk} {{2 \pi \alpha' \mu} \over 2} \left( J^{-ij\left(k\right)} + {\cal{O}}(\alpha'^{9 \over 2})\right) \Bigr] 
\end{eqnarray*}
where $\partial_A = {\partial \over {\partial U^A}}$ and 
$F^2 = \left({2 \pi \alpha' \mu} \over 3\right)^2 \left(\left(U^1\right)^2 + \cdots \left(U^{3}\right)^2 \right) 
+ \left({2 \pi \alpha' \mu} \over 6\right)^2 \left(\left(U^{4}\right)^2 + \cdots \left(U^{9}\right)^2 \right)$. 
For details of this calculation, see appendix. 
And $T^{--\left(AB\right)} \sim {\cal{O}}(\alpha'^{7 \over 2})$ and $J^{-ij\left(k\right)} \sim {\cal{O}}(\alpha'^{5 \over 2})$ are given by
\[
T^{--} = {\left(2 \pi \alpha'\right)^4 \over 4 g_s l_s} STr\left( F_{ab} F_{bc} F_{cd} F_{da} - {1 \over 4} F_{ab} F_{ab} F_{cd} F_{cd} 
+ \mbox{($\Psi^2$, $\Psi^4$ terms)} \right)
\]
\[
J^{-ij} = {\left(2 \pi \alpha'\right)^3 \over 6 g_s l_s} STr\left(\dot{U^i} \dot{U^l} F_{lj} - \dot{U^j} \dot{U^l} F_{li} 
- {1 \over 2} \dot{U^l} \dot{U^l} F_{ij} + {1 \over 4} F_{ij} F_{lm} F_{lm} + F_{im} F_{ml} F_{lj} + \mbox{($\Psi^2$, $\Psi^4$ terms)} \right)
\]
where $F_{0i} = \dot{U^i}$, $F_{ij} = i \left[U^i,U^j\right]$ and $a, b, \cdots$ run from 0 to 9, but $i, j, \cdots$ run 1 to 9.
More details of the fermion terms are given in page 17 of \cite{sugracurrent}. The higher multipole moments of these currents are given by
\[
T^{--\left(AB\right)} = Sym\left(T^{--}; U^A U^B\right) + T^{--\left(AB\right)}_{fermion}
\]
\[
J^{-ij\left(k\right)} = Sym\left(J^{-ij}; U^k \right) + J^{-ij\left(k\right)}_{fermion}
\]

In the decoupling limit (\ref{decouplinglimit}), 
\begin{eqnarray}
S_{flat} \to {4 \pi^2 l_s^3 \over g_s}\int dt 
Tr \left(
\sum_{i=1}^9 {1 \over 2} \dot{U}^{i2} 
+ {1 \over 4} \sum_{i,j=1}^9 \left[U^i , U^j \right]^2 
+ i \Psi^T \dot{\Psi} 
- \sum_{i=1}^9 \Psi^T \gamma^i \left[ U^i , \Psi \right] 
\right)
\end{eqnarray}
and 
\begin{eqnarray}
S_{linear} 
&\to&  
\int dt \Bigl[ - {1 \over 16} \partial_A \partial_B F^2 \left( T^{--\left(AB\right)} \right) 
+ \epsilon_{ijk} {2 \pi \alpha' \mu \over 2} J^{-ij\left(k\right)} \Bigr] \nonumber \\
&=& {4 \pi^2 l_s^3 \over g_s}\int dt \nonumber \\
& & Tr \Bigl[ 
-{1 \over 32} \left({{4 \pi^2 \alpha'^2 \mu} \over 3} \right)^2 
\sum_{A=1}^3 
STr \Bigl(Sym\bigl(F_{ab} F_{bc} F_{cd} F_{da} - {1 \over 4} F_{ab} F_{ab} F_{cd} F_{cd} \bigr) ;U^A U^A \Bigr) \nonumber \\
& &  -{1 \over 32} \left({{4 \pi^2 \alpha'^2 \mu}\over 6}\right)^2 \sum_{A=4}^9 
STr \Bigl(Sym\bigl(F_{ab} F_{bc} F_{cd} F_{da} - {1 \over 4} F_{ab} F_{ab} F_{cd} F_{cd} \bigr) ;U^A U^A \Bigr) \nonumber \\
& & +\sum_{i,j,k=1}^3 \epsilon_{ijk} 
\left({{4 \pi^2 \alpha'^2 \mu} \over 12}\right) STr \Big(Sym\big(\dot{U^i} \dot{U^l} F_{lj} - \dot{U^j} \dot{U^l} F_{li} \nonumber \\
& &- {1 \over 2} \dot{U^l} \dot{U^l} F_{ij} + {1 \over 4} F_{ij} F_{lm} F_{lm} + F_{im} F_{ml} F_{lj} \bigr);U^k \Bigr) \nonumber \\
& & +\mbox{($\Psi^2$, $\Psi^4$ terms)} \Bigr] 
\end{eqnarray}  

We conjecture that this is the dual field theory action. 
Note that the exact cancellation of 
the mass terms (which are ${\cal{O}}(\mu^2 U^2)$) in this theory is crucial because they are not well-defined in the decoupling limit (\ref{decouplinglimit}). 
The first nontrivial effects of the background (\ref{antiDSRmetric}) 
appears as higher ${\cal{O}}(\alpha'^4 \mu^2  U^{10})$, ${\cal{O}}(\alpha'^4 \mu^2  \dot{U}^{2} U^6)$ and ${\cal{O}}(\alpha'^4 \mu^2  \dot{U}^{4} U^2)$ terms which are well-defined. 
The symmetry breaking $SO(9) \to SO(3) \times SO(6)$ because of $\mu \alpha'^2$ terms 
are easily confirmed just as in the dual geometry (\ref{g_0def}), (\ref{decouplingmetric}). 
Also note that the above theory has 
classical vacuum where all $U^i$ take static diagonal form. 
The existence of Higgs branch in dual field theory is expected,
since the D0-branes can move freely in the background (\ref{antiDSRmetric}),  
as easily understood from 11-dimensional solution (\ref{H}). 

Let's study when the perturbation theory is valid in this field theory. 
The dimensionless effective coupling constants of this theory are 
$g_{eff}^2 = {g_{YM}^2 N \over U^3} \left( {\alpha'^2 \mu} U^3 \right)^a $ 
where $a$ is some 
constant which takes $0 \le a \le 1$. 
Therefore for perturbation theory to be valid, it is necessary to require 
following two conditions:
\be
\label{pertcondition}
\left( {\alpha'^2 \mu}\right) \left( g_{YM}^2 N\right) \ll 1 \hspace{2mm} \mbox{and} \hspace{2mm} {g_{YM}^2 N \over U^3} \ll 1
\ee
Note that this is the parameter region where the curvature of the dual geometry becomes large, see (\ref{U_mu>U_gy}).  
This is expected result of gauge/string duality as in the $\mu = 0$ case in \cite{IMSY}. 
Here we base our the argument on the bosonic terms, 
but the fermion terms don't change the above arguments at all. 
Finally we comment on the validity of weak approximation. 
From the geometry (\ref{g_0def}), 
it is reasonable to expect that weak approximation is valid as long as $U \ll \left( g_{YM}^2 N\right)^{1 \over 9} \left( {\alpha'^2 \mu}\right)^{- {2 \over 9}}$.

%%%%%%%%%%%%%%%%%%%%%%%%%%%%%%%%%%%%%%%%%%%%%%%%%%%%%%%%%%%%%%%%%%%%%%%%%%%%%%
\section{M(atrix) model on general pp-wave}
%%%%%%%%%%%%%%%%%%%%%%%%%%%%%%%%%%%%%%%%%%%%%%%%%%%%%%%%%%%%%%%%%%%%%%%%%%%%%%
In this section, we study less supersymmetric pp-wave backgrounds. One of the surprising 
things in pp-wave is that 
there are plenty of pp-wave solutions which preserve 
``supernumerary'' or fractional number of additional supersymmetries. 
So far as we know 
pp-wave solutions preserve ${\cal{N}}= 16 + {\cal{N}}_{extra}$, where 
${\cal{N}}_{extra} = 0,2,4,6,8,10$ \cite{GH1,CLP1,Michelson1,CLP2,GH2,
Michelson2}. The goal of this section is to try to derive the 
DLCQ action of the Matrix model on these general pp-waves. We summarize the 
relevant known results about supersymmetry of fractional pp-waves in the 
appendix.

The 11 dimensional pp-wave we consider has metric \cite{exactpp},
\[
ds^2 = 2 dx^{+} dx^{-} + H\left( x^i, x^{+} \right) dx^{+2} + 
\sum_{i=1}^9 dx^{i2}
\]
\[
H\left( x^i, x^{+} \right) = -\sum_{i=1}^9 \mu_i^2 x^{i2}
\]
\[
F^{\left(4 \right)} = \sum_{i,j,k=1}^9 {1 \over 3!} 
f_{ijk} dx^{+} \wedge dx^{ijk}
\]
The Einstein equations require,
\[
\sum_{i=1}^9 \mu_i^2 = {1 \over 2}\left( {1 \over 3!} \right) 
\sum_{i,j,k=1}^9 \left(  f_{ijk} \right)^2
\]
Here pp-waves always have at least 16 supercharges, plus $2k$ ``extra'' 
supersymmetries (see eq. (\ref{chi}),(\ref{constraint}),(\ref{det}),(\ref{susycondition}) in the 
appendix).

%%%%%%%%%%%%%%%%%%%%%%%%%%%%%%%%%%%%%%%%%%%%%%%%%%%%%%%%%%%%%%%%%%%%%%%%%%%%%%
\subsection{Supermembrane on general pp-wave}
%%%%%%%%%%%%%%%%%%%%%%%%%%%%%%%%%%%%%%%%%%%%%%%%%%%%%%%%%%%%%%%%%%%%%%%%%%%%%%
The 11 dimensional supermembrane action is 
\[
S \left[Z \left( \zeta \right) \right] 
= \int d^3 \zeta \left[ - \sqrt{ - G\left(Z\right)} 
- {1 \over 6} \epsilon^{abc} \Pi^A_a \Pi^B_b \Pi^C_c B_{CBA} 
\left(Z\left(\zeta\right) \right) \right]
\]
where $Z^A = \left( X^{M}\left(\zeta \right),\xi\left( \zeta \right) \right)$ 
are superspace embedding coordinates and $B_{CBA}$ the antisymmetric 
tensor gauge superfield, 
\footnote{We use notation $A = \left( M , \alpha \right)$ for curved space-time indices, here $M =\left( +,-, i \left(=1,...,9 \right) \right)$ whereas tangent space vector indices $r = +, -, i \left(=1,...,9\right)$. 
We are sloppy about transverse coordinates because that direction is flat. 
Also we use $a=0,1,2$ for world-volume coordinate on membrane such that $\epsilon_{012} = 1$.} 
we also take light cone gauge $X^+ = \tau$, 
$\tau$ is the world-volume time coordinate \footnote{We choose world-volume coordinates $\left( \tau, \sigma^1, \sigma^2 \right)$.}, 
and $G$ is given by
\[
G\left( X,\xi \right) = det \left( \Pi^M_a \Pi^N_b g_{MN} \right) = 
det \left( \Pi^r_a \Pi^s_b \eta_{rs} \right).
\]
Here $\Pi^r_a$ is the supervielbein pullback
\[
\Pi_a^r = \partial_{a} Z^A E_A^r 
\]

The supermembrane action on fully supersymmetric pp-wave 
background can be derived by using 
the coset space approach where we can express the supervielbein and tensor 
gauge superfield background in all orders of $\xi$ \cite{WitPPS}. 
In the fractional 
pp-wave case, we don't know the exact supervielbein and 
tensor field, although it is known to order $\left( \xi \right)^2$ 
\cite{WitPP}. In the case that the gravitino has 0 vev, 
supervielbein pullback $\Pi^r_a$ is given by
\begin{eqnarray*}
\Pi_a^r &=& \partial_{a} Z^A E_A^r \\
&=& \partial_{a} X^M \left( e_M^r
-{1 \over 4} \bar{\xi} \Gamma^{rst} \xi w_{Mst} 
+ \bar{\xi} \Gamma^r \Omega_M \xi \right) + \bar{\xi} \Gamma^r \partial_{a} 
\xi + {\cal{O}}\left(\xi^4 \right)
\end{eqnarray*}
Also the tensor fields pullback term is given by
\begin{eqnarray*}
- {1 \over 6} \epsilon^{abc} \Pi^A_a \Pi^B_b \Pi^C_c B_{CBA} 
\left(Z\left(\zeta\right) \right) 
= {1 \over 6} \epsilon^{abc} \partial_a X^M \partial_b X^N \partial_c X^L 
\left[ C_{MNL} + {3 \over 4} \bar{\xi} \Gamma_{rs} \Gamma_{MN} \xi w_L^{rs} 
- 3 \bar{\xi} \Gamma_{MN} \Omega_L \xi \right] 
\\
- \epsilon^{abc} \bar{\xi} \Gamma_{MN} \partial_c \xi \left[ {1 \over 2} 
\partial_a X^M \left( \partial_b X^N + \bar{\xi} \Gamma^N \partial_b \xi 
\right) + {1 \over 6} \bar{\xi} \Gamma^M \partial_a \xi \bar{\xi} \Gamma^N 
\partial_b \xi \right] + {\cal{O}}\left(\xi^4 \right)
\end{eqnarray*}
Here the nonzero components of $w_M^{rs}, \Omega_M$ are given in 
eq. (\ref{nonzerocomponent}) in the appendix.
In order to write down the explicit form of the fermion fields, we write the 
11 dimensional gamma matrices in terms of 9 dimensional gamma matrices as 
given in eq. (\ref{gamma}) and remove the $\kappa$ symmetry of this action by gauge fixing 
\[
\Gamma^{+} \xi = 0 
\]
\[
\Leftrightarrow\bar{\xi} = {1 \over2^{1 \over 4}} 
\pmatrix{
0 & -\Psi^{T}  \cr
} , \, \hspace{2mm} 
\xi \equiv {1 \over 2^{1 \over 4}} \pmatrix{
\Psi \cr
0 \cr
}
\]
Under this gauge condition, the supervielbein pullback and tensor gauge 
superfield pullback are written as 
\[
\Pi^{+}_a = \partial_a X^+, \hspace{3mm}  \, \Pi^i_a = \partial_{a} X^i, 
\]
\[
\Pi^{-}_a = \partial_a X^-  + \partial_a X^+  \left( {H \over 2} - {i \over 4} \Psi^{T} \theta \Psi \right) 
+ i \Psi^{T} \partial_{a} \Psi, 
\]
\[
- {1 \over 6} \epsilon^{abc} \Pi^A_a \Pi^B_b \Pi^C_c B_{CBA} 
= \left(- {1 \over 6} \sum_{i,j,k=1}^9 f_{ijk} \{X^i , X^j \}_{PB} X^k 
-i \sum_{i=1}^9 \Psi^T \gamma^i \{ X^i , \Psi \}_{PB} 
\right)
\]
Here $\{X,\Psi\}_{PB} = \epsilon^{0ab} \partial_a X \partial_b \Psi$, and  
$\theta \equiv {1 \over 3!} f_{ijk} \gamma^{ijk}$ is the 16 $\times$ 16 
$SO(9)$ gamma matrix. Therefore the action is given by 
\begin{eqnarray}
\label{supermembraneaction}
S = \int d^3 \sigma 
\Biggl(
\sum_{i=1}^9 {1 \over 2} \dot{X}^{i2} 
- \sum_{i=1}^9 {1 \over 2} \mu_i^2 X^{i2} 
- {1 \over 4} \sum_{i,j=1}^9 \{X^i , X^j \}^2_{PB} 
- {1 \over 6} \sum_{i,j,k=1}^9 f_{ijk} \{X^i , X^j \}_{PB} X^k 
\nonumber \\
+ i \Psi^T \dot{\Psi} 
- {i \over 4} \Psi^T \theta \Psi 
-i \sum_{i=1}^9 \Psi^T \gamma^i \{ X^i , \Psi \}_{PB} 
\Biggr)
\end{eqnarray}
again up to ${\cal{O}}\left( \Psi \right)^4$. 

So far we have derived this action by truncating the full action, and  
in principle we should include terms of higher order in $\Psi$. But
even though we derived this action by the truncation of  
${\cal{O}}\left(\Psi \right)^4$ terms, we can argue 
that this is the exact action 
on general pp-wave background. The proof is as follows. 
\footnote{Similar argument about Green-Schwarz action 
on general pp-wave background which doesn't have supernumerary supersymmetry 
is found in \cite{RussoTseytlin}.}  
The truncated terms have two parts, one being higher order terms in the 
supervielbein pullback, the other higher order terms in 
the tensor gauge superfield pullback. 
Let's consider the supervielbein first. 
The supervielbein is defined as $\Pi^r_i = \partial_i Z^A E^r_A$, and 
$\partial_i Z^A$ is proportional to $\partial_i X^M$ or $\partial_i \xi$, 
while  
$E^r_A$ is superfield of elfbein which is made by 
some complicated function of the quantities $\xi$, $\Gamma^r$, $C_{ijk}$, 
$F_{+ijk}$, $\partial_M$, $w_{MNL} \equiv e_{rN}e_{sL} w_M^{rs}$, 
$\Gamma^M_{NL}$ and all other geometrical functions with indices of 
curved space $M,N, ...$, i.e. (some derivative of
 $e^r_M$) $\times$ $e_{rN}$. 
But note that $C_{ijk}$, $F_{+ijk}$, $\partial_M$, 
$w_{MNL} \equiv e_{rN} e_{sL} w_M^{rs}$, $\Gamma^M_{NL}$ and 
whatever function 
we obtain from the derivative of elfbein, they cannot have curved space-time 
lower index $M=-$. This is because the pp-wave 
has null Killing vector $k^M$, which has only one nonzero component 
along upper index $M=-$ and lower index $M=+$, and the field strength 
is proportional to $k_M = k_{+}$. 
In the same way all upper curved space indices 
can't have $M=+$. 
On the other hand, because of the gauge condition $\Gamma^+ \xi
 = 0$, $\bar{\xi} \Gamma^r ... \Gamma^s \xi$ can be nonzero
 if and only if we have 
one tangent space upper index $r={-}$ and no upper index $r={+}$ 
for the Gamma matrices between $\bar{\xi}$ and $\xi$. 
Since $e_{r={-},M} \ne 0$ if and only if $M=+$, 
in order to contract the curved 
space-time index $M$, we need upper $M = {+} $ 
for each $\bar{\xi} \Gamma^r ... \Gamma^s \xi$ term, and 
the only component which can have upper $M=+$ 
to contract is $\partial_a X^{\left( M= {+} \right) } = \delta_a^0$. 
Since the supervielbein pullback is at most linear 
in $\partial_a X^M$, we conclude that $\xi$ is at most 
bilinear in $\xi$, therefore that 
there are no higher order corrections in $\xi$. 

In the same way we can show that there are no higher order corrections to 
the $\epsilon^{abc} \Pi^A_a \Pi^B_b \Pi^C_c B_{CBA} 
\left(Z \left(\xi \right) \right)$ term in the action.  
Since each bilinear term $\bar{\xi} \Gamma^r ... \Gamma^s \xi$ needs 
one upper index $r={-}$ and $e_{r={-}, M} \ne 0$ iff $M= {+}$, 
we need $\partial_a X^{\left( M= {+} \right) } = \delta_a^0$ from 
$\partial_a Z^A$ . 
But because $\epsilon^{abc} \partial_a X^{M= {+}} \partial_b X^{N={+}} = 0$, 
we can have at most linear terms in $\partial_i X^{M= {+}} $. 
Therefore these terms are at most bilinear in $\xi$ 
so the action (\ref{supermembraneaction}) which we obtained is exact. 

%%%%%%%%%%%%%%%%%%%%%%%%%%%%%%%%%%%%%%%%%%%%%%%%%%%%%%%%%%%%%%%%%%%%%%%%%%%%%%
\subsection{Matrix model and its supersymmetry}
%%%%%%%%%%%%%%%%%%%%%%%%%%%%%%%%%%%%%%%%%%%%%%%%%%%%%%%%%%%%%%%%%%%%%%%%%%%%%%
Once we obtain the supermembrane action exactly, it is straightforward to 
derive the Matrix model action by regularizing according to the usual prescription \cite{deWitHoppeNicolai}.
\[X^M \left(\zeta \right) \leftrightarrow X^M_{N \times N}\]
\[\Psi \left(\zeta \right) \leftrightarrow \Psi_{N \times N} \]
\[
\int d^2 \sigma \leftrightarrow Tr 
\]
\[
\{ , \}_{PB} \leftrightarrow - i \left[ ,  \right] 
\]
\begin{eqnarray}
S = \int d\tau 
Tr \Biggl(
\sum_{i=1}^9 {1 \over 2} \dot{X}^{i2} 
- \sum_{i=1}^9 {1 \over 2} \mu_i^2 X^{i2} 
+ {1 \over 4} \sum_{i,j=1}^9 \left[X^i , X^j \right]^2 
+ {i \over 3} \sum_{i,j,k=1}^9 f_{ijk} X^i X^j X^k 
\nonumber \\
+ i \Psi^T \dot{\Psi} 
- {i \over 4} \Psi^T \theta \Psi 
- \sum_{i=1}^9 \Psi^T \gamma^i \left[ X^i , \Psi \right] 
\Biggr)
\end{eqnarray}

Note that on general pp-wave, the mass term is time dependent. 
Remember that in BFSS Matrix model description of M-theory, 
we have 16 dynamical supersymmetries of the action, 
and that the supersymmetry transformation spinor is the 
one for the background which survives after we introduce 
momentum. 
That is, the 16 components of 
the background Killing spinor which 
satisfy $\Gamma_{01} \epsilon = \epsilon \left( \Leftrightarrow 
\Gamma^{-} \epsilon = 0 \right)$ are realized as the dynamical 
supersymmetries of the flat space Matrix theory. 
The same story should hold for a Matrix model on a pp-wave background. 
Actually one can see that the 
dynamical supersymmetry transformation spinor of 
BMN action has exactly this form as given in eq. (\ref{chi}) $\chi_{-}$ 
in the appendix. (Note that because $\Gamma^{+} \epsilon = \Gamma^{+} \chi 
= 0 \Leftrightarrow \chi_{-} = 0$, 
the $\chi_{+}$ components are not realized as dynamical supersymmetry, but 
rather realized as kinematical supersymmetry of the Matrix model on pp-wave.)
In the same way we expect the above action has $2k$ supersymmetries  
as expected from the background pp-wave geometry.
Now it is straightforward to check the 
condition of supersymmetry of this action. In order to analyze the 
supersymmetry of 
 this action, it is enough to analyze just the Abelian case. We therefore 
consider the Abelian action
\be
\label{particleaction}
S = \int d \tau \left(\sum_{i=1}^9 {1 \over 2} \dot{x}^{i2}
 - \sum_{i=1}^9 \mu_i^2 x^{i2} + \Psi^{T} \partial_{\tau} \Psi 
 - { a \over 4} \Psi^{T} \theta \Psi \right) \,.
\ee

Even though we expect $a = 1$ from an derivation, here we take $a$ to be  
an unknown constant. Taking the ansatz of supersymmetry transformation as,
\[
\delta X^i = \Psi^{T} \gamma^i {\cal{N}}_i \varepsilon
\left( \mbox{no sum over $i$} \right)
\]
\[
\delta \Psi = \sum_{i=1}^9 \left( \dot{X}^i \gamma^i {\cal{N}}_i \varepsilon 
+ \mu X^i \gamma^i {\cal{M'}}_i \varepsilon \right)
\]
\[
\varepsilon =  \varepsilon \left( \tau \right) 
= e^{\mu M \tau} \varepsilon_0
\]

We wish to determine the matrices ${\cal{N}}_i$, ${\cal{M'}}_i$, $M$ 
as well as $a$.
Terms of ${\cal{O}}\left(\mu^0\right)$ automatically cancel. 
Terms of ${\cal{O}}\left(\mu^1\right)$ give
\be
\label{mu1}
{\mu \cal{M'}}_i = - \mu {\cal{N}}_i M + {a \over 4} \theta^i {\cal{N}}_i 
\ee
and 
\[
\dot{\mu} = 0.
\]
Here $\theta^i \equiv \gamma^i \theta \gamma^i$. Finally the 
${\cal{O}}\left(\mu^2\right)$ terms give 
\be
\label{mu2}
\left(
\mu^2 {\cal{N}}_i M^2 - {a \mu \over 2} \theta^i {\cal{N}}_i M 
+\left( \left( {a \over 4}\right)^2  \theta^{i2}  
+ \mu_i^2 \right)
 {\cal{N}}_i \right) \varepsilon = 0 
\ee
for all $i=1 ... 9$. Let's consider some solutions of this 
equation. Setting the matrices ${\cal{N}}_i =1$ for all 
$i=1...9$, equation (\ref{mu2}) becomes 
\[
\left(
\mu^2 M^2 - {a \mu \over 2} \theta^i M 
+  \left( {a \over 4}\right)^2  \theta^{i2}  + \mu_i^2 \right)
 \varepsilon = 0.
\]
By setting $M \equiv b {\theta \over \mu}$, with $b$ some unknown constant, this equation 
reduces to 
\[
\left( b^2 \theta^2 -  {a b \over 2} \theta^i \theta 
+ \left( {a \over 4}\right)^2 \theta^{i2}
+ \mu_i^2 \right) \varepsilon = 0.
\]

You can see that the solution $a = \pm 1$, $b = \mp {1 \over 12}$ exactly coincides 
with eq. (\ref{constraint}). Also the form of the spinor is given by 
$ \varepsilon = e^{\mp {\mu \over 12} \theta}$ and again this is exactly 
the expected form of the $2k$ Killing spinor in (\ref{chi}).
So far we have found that some supersymmetry transformations of the 
Matrix action are inheritated from the background action.

Note also that this action has kinematical ${\cal{N}} = 16 $ 
supersymmetry under which fields transform as:
\[
\delta X^i = 0
\]
\[
\delta \Psi = exp \left(+ 
{1 \over 4} \int^{x^+} d{x'^+} \theta \left( {x'^+} \right) \right) \eta
\]
for some constant spinor $ \eta $. This is again expected from 
the supersymmetry of the background.

%%%%%%%%%%%%%%%%%%%%%%%%%%%%%%%%%%%%%%%%%%%%%%%%%%%%%%%%%%%%%%%%%%%%%%%%%%%%%%
\section{Conclusions}
%%%%%%%%%%%%%%%%%%%%%%%%%%%%%%%%%%%%%%%%%%%%%%%%%%%%%%%%%%%%%%%%%%%%%%%%%%%%%%
In the first part, we have studied the back-reaction of the large-N momentum on the 
maximally supersymmetric anti-pp-wave geometry. By taking an appropriate decoupling 
limit, we derived the type IIA geometry on which string theory is well-defined. 
The field theory action which we conjecture to be dual to the geometry is 
derived through the linear coupling between D0-branes and background fields. 
This gives a deformed version of 0+1-dimensional SYM 
with 16 supercharges, dual to warped $AdS_2 \times S^8$ geometry 
deformed by ${\cal{O}}\left(\mu \alpha'^2\right)$ terms. 
Both theories possess $SO(3) \times SO(6)$ symmetry and Higgs branch. Also 
similarly perturbation theory and dual geometry don't make sense at the same time. 
In the second part, we succeed in deriving the supermembrane action and the Matrix model 
on a general pp-wave background only through the properties of null Killing vector 
and checked that the supersymmetry 
of this action is as expected from the background. There are several 
points which are not fully understood. It would be nice to understand the 
amount of supersymmetry in the geometry (\ref{decouplingmetric}), and
also it would be nice to understand whether there are any other vacua in the 
dual field theory. And of course it is very nice to 
derive the 11-dimensional supergravity solution of D0-branes on usual 
pp-wave, if it exists. Finally it would be interesting to analyze the light cone time-dependent 
Matrix theory which is dual to M-theory on a time dependent pp-wave.   

\bigskip
\goodbreak
\centerline{\bf Acknowledgments}
\noindent
The author is very grateful to Mark Jackson, Koenraad Schalm 
and especially Dan Kabat for helpful comments, discussion and encouragement. 
In the first version of this paper, 
the author made a mis-interpretation of anti-D0-branes as D0-branes 
and therefore lead to misunderstanding of dual field theory. 
The author want to thank specially Juan Maldacena 
for his prompt correspondence and helpful comments about dual field theory interpretation.
This research is supported by DOE under contract DE-FG02-92ER40699.

\appendix
%%%%%%%%%%%%%%%%%%%%%%%%%%%%%%%%%%%%%%%%%%%%%%%%%%%%%%%%%%%%%%%%%%%%%%%%%%%%%%
\section{Fractionally supersymmetric pp-waves}
%%%%%%%%%%%%%%%%%%%%%%%%%%%%%%%%%%%%%%%%%%%%%%%%%%%%%%%%%%%%%%%%%%%%%%%%%%%%%%
Here we summarize the known results of 11 dimensional general 
pp-waves which posses $16 \le {\cal{N}} \le 32$ SUSY. 

The 11 dimensional pp-wave has metric
\[
ds^2 = 2 dx^{+} dx^{-} + H\left( x^i, x^{+} \right) dx^{+2} + 
\sum_{i=1}^9 dx^{i2}
\]
\[
H\left( x^i, x^{+} \right) = -\sum_{i=1}^9 \mu_i^2 x^{i2}
\]
\[
F^{\left(4 \right)} = \sum_{i,j,k=1}^9 {1 \over 3!} 
f_{ijk} dx^{+} \wedge dx^{ijk}
\]
\[
\sum_{i=1}^9 \mu_i^2 = {1 \over 2 \cdot 3!} 
\sum_{i,j,k=1}^9 \left(  f_{ijk} \right)^2
\]
Generally $\mu_i$ and $f_{ijk}$ are functions of $x^{+}$.

The Killing spinor $\epsilon$ should satisfy
\be
\label{Killingeq}
{\cal{D}}_{M} \epsilon = \nabla_{M} \epsilon  + \Omega_{M} \epsilon = 0 \,.
\ee
\[
\Omega_{M} = {1 \over 288} \left( \Gamma_{M}^{PQRS} - 8 \delta^P_M \Gamma^{QRS} 
\right) F_{PQRS} 
\]
for $M= 0,1, ... ,11$.
In a pp wave background, the non-zero components of the elfbein $e_{rM}$, 
spin connection $w_M^{rs}$ and 
$\Omega_{M}$ are
\be
\label{nonzerocomponent}
e_{r=+,M=+} = {H \over 2} ,\, e_{r=+,M=-} = 1 ,\,
e_{r=-,M=+} = 1 ,\, e_{r=i,M=j} = \delta_{ij} \,
\ee
\[
w_{+}^{-i} = {1 \over 2} \partial_i H = \mu_i^2 x^i\,.
\]
\[
\Omega_{+} = - {1 \over 12} \Theta \left(\Gamma_{-} \Gamma_{+} +1 \right)
\]
\[
\Omega_i = {1 \over 24} \left(3 \Theta \Gamma_i + \Gamma_i \Theta \right) 
\Gamma_{-}
\]
where $\Theta$ is defined as 
\[
\Theta \equiv \sum_{ijk=1}^9 {1 \over 3!} f_{ijk} \Gamma^{ijk}.
\]

In order to discuss the supersymmetry of these pp-wave backgrounds, 
it is convenient to decompose the 32-component Killing spinor 
$\epsilon$ in terms of two 16-component $SO(9)$ spinors as
\[
\epsilon = \sum_{i=1}^9 \left(1 - x^i \Omega_i \right) \chi, \hspace{3mm}
\chi \equiv 
\pmatrix{
\chi_{+} \cr
\chi_{-} \cr
}
\]
Here we extract $x^i$ dependence from $\epsilon$ explicitly, 
therefore $\chi \equiv 
\pmatrix{
\chi_{+} \cr
\chi_{-} \cr
}$ depends only on $x^{+}$. One can check that this $x^i$ dependence 
satisfies (\ref{Killingeq}). 

Decomposing the 11 dimensional gamma matrices in terms of $SO(9)$ gamma 
matrices, 
\be
\label{gamma}
\Gamma^{i} = \gamma^i \times \sigma_3  \left(i = 1...9 \right) ,\hspace{2mm} \,
\Gamma^{0} = 1 \times i \sigma_1 ,\hspace{2mm} \,
\Gamma^{11} = -1 \times  \sigma_2 
\ee
\[
\Gamma_{-} = \Gamma^{+} =
{1 \over \sqrt{2}} \left(\Gamma^0 + \Gamma^{11} \right) 
= \sqrt{2} 
\pmatrix{
0 & i \bf{1} \cr
0 & 0 \cr
} 
\]
\[
 \Gamma_{+} = \Gamma^{-} 
=  {1 \over \sqrt{2}} \left(-\Gamma^0 + \Gamma^{11} \right)    
= \sqrt{2}
\pmatrix{
0 & 0 \cr
-i \bf{1} & 0 \cr
} 
\]
and define the 9 dimensional gamma matrix $\theta$ 
\[
\Theta \equiv \sum_{i,j,k=1}^9 {1 \over 3!} f_{ijk} \Gamma^{ijk}
= \sum_{i,j,k=1}^9 {1 \over 3!} f_{ijk} \gamma^{ijk} \times \sigma_3
\equiv \theta \times \sigma_3 \,.
\]
Therefore $\Gamma^+ \epsilon = 0$ means $\chi_{-} = 0$ 
since $\Gamma^{+} \epsilon = \Gamma^{+} \chi$.
Then the Killing spinor which satisfies (\ref{Killingeq}) is expressed as 
\be
\label{chi}
\chi_{+} = exp \left(+
{1 \over 4} \int^{x^{+}} dx'^+ \theta \right) \psi_{+} ,\, \hspace{2mm} 
\chi_{-} = exp \left(- {1 \over 12} \int^{x^{+}} dx'^+ \theta \right) \psi_{-} \,.
\ee
Here $\psi_{\pm}$ is a constant spinor. Equation (\ref{Killingeq}) gives 
no constraint on $\psi_+$ but gives a constraint on $\psi_{-}$, that 
\be
\label{constraint}
\left[ 144 \mu_i^{2} + \left(3 \gamma^i \theta \gamma^i + \theta \right)^2 
-12 \left(3 \gamma^i \partial_{{+}}\theta \gamma^i 
+ \partial_{{+}}\theta \right)
\right] \psi_{-} = 0 
\ee
for all $i = 1...9$. 
Here $\partial_{+} = {{\partial} \over {\partial x^{+}}}$. 
The condition that there exist supernumerary supersymmetry is 
that there exists non-trivial $\psi_{-}$ which satisfy this equation, 
that is, the determinant must be zero, or
\be
\label{det}
\Pi_{I=1}^{8} \{ \left( 
\left(12 \mu_i \right)^2 - \rho_{i,I}^2 \right)^2 + \left(12 \partial_{{+}} \rho_{i,I} \right)^2  \}  = 0 
\ee
for all $i = 1...9$. 
Here we choose the basis so that the 16 $\times$ 16 matrix 
$\left(3 \gamma^i \theta \gamma^i + \theta \right)$ is skew diagonal 
and its skew eigenvalues are $\rho_{i,I}$ where $I = 1,...,8$.

This shows that $f_{ijk}$ should be $x^{+}$-independent in order to have 
supernumerary supersymmetry. Then (\ref{det}) can be zero if
\be
\label{susycondition}
144 \mu_i^2 = \rho_{I_1 i}^2 = \cdots 
= \rho_{I_k,i}^2 
\ee
is satisfied for all $i=1 ... 9$.
Therefore pp-wave backgrounds always possess 16 supersymmetries 
from $\psi_+$, 
and $2k$ supersymmetries 
from $\psi_{-}$ when the background has no $x^{+}$-dependence. 
$\psi_{-}$ has (16-2k) zero components 
because of the $16-2k$ nontrivial constraints (\ref{constraint}).

%%%%%%%%%%%%%%%%%%%%%%%%%%%%%%%%%%%%%%%%%%%%%%%%%%%%%%%%%%%%%%%%%%%%%%%%%%%%%%
\section{Dual field theory action in detail}
%%%%%%%%%%%%%%%%%%%%%%%%%%%%%%%%%%%%%%%%%%%%%%%%%%%%%%%%%%%%%%%%%%%%%%%%%%%%%%
The dual field theory has action:
\[
S = S_{flat} + S_{linear}
\]
\begin{eqnarray*}
S_{linear} &=& \int dt \Bigl[ {1 \over 4} \partial_A \partial_B  h_{\mu \nu} I^{\mu \nu \left( A B \right)}_h 
+ {1 \over 2} \partial_A \partial_B \phi I^{\left( A B \right)}_{\phi} \\
& & + {1 \over 2} \partial_A \partial_B 
A_{\mu} I^{\mu \left( A B \right)}_{0} 
+ \partial_A  B_{\mu \nu} I^{\mu \nu \left( A \right)}_{s} + \partial_A  C_{\mu \nu \lambda} 
I^{\mu \nu \lambda \left( A \right)}_{2} \Bigr] \\
&=& \int dt \Bigl[ -{1 \over 32} \partial_A \partial_B F^2 \left( I^{00 \left(AB\right)}_h + I^{ii \left(AB\right)}_h \right) 
- {3 \over 32} \partial_A \partial_B F^2 I^{\left(AB\right)}_{\phi} \\
& & + {1 \over 8} \partial_A \partial_B F^2 I^{0 \left( A B \right)}_{0} 
- \epsilon_{ijk} {{2 \pi \alpha' \mu} \over 6} I^{ij \left( k \right)}_{s} 
+ 3  \epsilon_{ijk} {{2 \pi \alpha' \mu} \over 6} I^{0ij \left( k \right)}_{2} \Bigr] \\
&=& \int dt \Bigl[ -{1 \over 32} \partial_A \partial_B F^2 
\left( T^{++\left(AB\right)} + T^{+-\left(AB\right)} + (I^{00\left(AB\right)}_h)_8 
+ T^{ii\left(AB\right)} + (T^{ii\left(AB\right)}_h)_8 +{\cal{O}}(\alpha'^{11 \over 2}) \right) \\
& & - {3 \over 32} \partial_A \partial_B F^2 \left(T^{++\left(AB\right)} -{1 \over 3} T^{+-\left(AB\right)} 
-{1 \over 3} T^{ii\left(AB\right)} + (I^{\left(AB\right)}_{\phi})_8 + {\cal{O}}(\alpha'^{11 \over 2}) \right) \\
& & + {1 \over 8} \partial_A \partial_B F^2 \left( T^{++\left(AB\right)} \right)
- \epsilon_{ijk} {{2 \pi \alpha' \mu} \over 6} \left(3 J^{+ij\left(k\right)} -3J^{-ij\left(k\right)} 
+ {\cal{O}}(\alpha'^{9 \over 2}) \right) \\
& & + 3 \epsilon_{ijk} {{2 \pi \alpha' \mu} \over 6} \left(J^{+ij\left(k\right)} 
+ {\cal{O}}(\alpha'^{9 \over 2})  \right) \Bigr] \\
&=& \int dt \Bigl[ - {1 \over 32} \partial_A \partial_B F^2 \left( (I^{00\left(AB\right)}_h)_8
 + (T^{ii\left(AB\right)}_h)_8 + 3 (I^{\left(AB\right)}_{\phi})_8 + {\cal{O}}(\alpha'^{11 \over 2}) \right) \\
& & + \epsilon_{ijk} {{2 \pi \alpha' \mu} \over 2} \left( J^{-ij\left(k\right)} + {\cal{O}}(\alpha'^{9 \over 2})\right) \Bigr] \\
&=& \int dt \Bigl[ - {1 \over 16} \partial_A \partial_B F^2 \left( T^{--\left(AB\right)} + {\cal{O}}(\alpha'^{11 \over 2})\right) 
+ \epsilon_{ijk} {{2 \pi \alpha' \mu} \over 2} \left( J^{-ij\left(k\right)} + {\cal{O}}(\alpha'^{9 \over 2})\right) \Bigr] 
\end{eqnarray*}
where $\partial_A = {\partial \over {\partial U^A}}$ and 
$F^2 = \left({2 \pi \alpha' \mu} \over 3\right)^2 \left(\left(U^1\right)^2 + \cdots \left(U^{3}\right)^2 \right) 
+ \left({2 \pi \alpha' \mu} \over 6\right)^2 \left(\left(U^{4}\right)^2 + \cdots \left(U^{9}\right)^2 \right)$. 
Note that terms involving $T^{++\left(AB\right)}$ and $J^{+ij\left(k\right)}$, which are origin of BMN mass term in pp-wave case, are exactly canceled out in this anti-pp-wave case.

%%%%%%%%%%%%%%%%%%%%%%%%%%%%%%%%%%%%%%%%%%%%%%%%%%%%%%%%%%%%%%%%%%%%%%%%%%%%%%%%

\end{document}